%% file: aa3647-05.tex
\documentclass[letterpaper]{aa}
\usepackage{hyperref}
\usepackage{natbib}
\usepackage{txfonts}
\usepackage{graphicx}
\bibpunct{(}{)}{;}{a}{}{,}

\renewcommand\pi\piup

\begin{document}

\title{Simple Models for the Distribution of Dark Matter}
\author{Jin H. An\thanks{Current address:
MIT Kavli Institute for Astrophysics \& Space Research,
Massachusetts Institute of Technology,
77 Massachusetts Avenue, Cambridge, MA 02139, USA}
\and N. Wyn Evans}
\institute{Institute of Astronomy, University of Cambridge,
Madingley Road, Cambridge CB3 0HA, UK;\\
\email{jinan@space.mit.edu, nwe@ast.cam.ac.uk}}
\date{{\it Astron.\ \& Astrophys.}\ {\bf 444}, 45-50 (2005)\hfill
{DOI: 10.1051/0004-6361:20053647}}

\abstract{
We introduce a simple family of models for representing the dark
matter in galaxies. The potential and phase space distribution
function are all elementary, while the density is cusped. The models
are all hypervirial, that is, the virial theorem holds locally, as
well as globally. As an application to dark matter studies, we
compute some of the properties of $\gamma$-ray sources caused by
neutralino self-annihilation in dark matter clumps.
}

\maketitle

\section{Introduction}

It is worthwhile to find simple models for the distribution of dark
matter in galaxy haloes. This is tantamount to solving the
collisionless Boltzmann and Poisson equations for the distribution
function (DF) $f$, potential $\psi$, and density $\rho$ of the dark matter
particles.

Many authors start by assuming a profile for the dark matter density
and then solving for the self-consistent potential and DF.
This has provided some widely-used and notable models for
dark matter haloes \citep[e.g.,][]{Ja83,He90}. A drawback to this
approach is that, even if the density and potential are simple, the
DF is often unwieldy. For example, the DF
of the \citet{Ja83} model is a higher transcendental
function, while the DF of the \citet{He90} model is
composed of an unwieldy bunch of elementary functions.

It can be advantageous to tackle this problem the other way around, by
assuming a simple DF and then solving for the
self-consistent potential and the density. Except for \citet{To82},
this reverse approach has not been widely used. Here, we exploit it to
build a flexible family of cusped dark matter halo models with an
elementary DF and potential (Sects.~\ref{sec:df} \& \ref{sec:shm}).

As a simple application, we use our models to study the signal from
indirect detection experiments (Sect.~\ref{sec:app}). In particular,
$\gamma$-rays from dark matter annihilation may be identified by
forthcoming atmospheric \v{C}erenkov telescopes such as
\textsl{VERITAS}\footnote{\url{http://veritas.sao.arizona.edu}} or by
satellite-borne detectors like
\textsl{GLAST}\footnote{\url{http://www-glast.stanford.edu}}, and so
it is useful to have definite predictions from halo models.

\section{A Simple distribution function}
\label{sec:df}

\subsection{An ansatz}

Let us assume that the dark halo is spherical, in which case the DF
may depend on the binding energy $E$ and the magnitude of the angular
momentum $L$. Let us note that the generalized Plummer models,
recently studied by \citet{EA05} have very simple power-law DFs of the
form
\begin{equation}
f(E,L)\propto L^{p-2}E^{(3p+1)/2}.
\end{equation}
This suggests that it may be worthwhile to look for models with
DFs given by the sum of such components
\citep[c.f.,][]{Fr52,To82};
\begin{equation}
f(E,L)=\sum_iC_iL^{p_i-2}E^{(3p_i+1)/2}.
\label{eq:ghv}
\end{equation}
where $C_i$ and $p_i$ are all constants. Then, the density is of the
form
\begin{equation}
\rho=\sum_iD_ir^{p_i-2}\psi^{2p_i+1},
\label{eq:den}
\end{equation}
where the constants $D_i$ are
\begin{equation}
D_i=2^{(p_i+1)/2}\pi^{3/2}C_i
\frac{\Gamma(p_i/2)\Gamma(3p_i/2+3/2)}{\Gamma(2p_i+2)}.
\label{eq:con1}
\end{equation}
By integrating the DF over velocity space, it is straightforward to
derive the radial and tangential velocity second moments $\langle
v_r^2\rangle$ and $\langle v_\mathrm{T}^2\rangle$ respectively.
Whereas the anisotropy parameter $\beta=1-\langle v_\mathrm{T}^2
\rangle/(2\langle v_r^2\rangle)$ is no longer constant in contrast to
models of \citet{EA05}, we still find the remarkably simple relation
between the three-dimensional velocity dispersion and the potential;
\begin{equation}
\langle v_r^2\rangle+\langle v_\mathrm{T}^2\rangle=\frac{\psi}{2}.
\label{eq:hyper}
\end{equation}
In other words, the root mean square velocity is always one-half of
the escape velocity at every spot, or the virial theorem holds locally
for any model described by the DF of Eq.~(\ref{eq:ghv}). \citet{EA05} coined
the term ``hypervirial'' to describe such stellar dynamical models for
which the kinetic energy in each volume element ($T=\rho\langle
v^2\rangle/2$) is exactly one-half of the magnitude of the local
contribution to the potential energy by the same volume element
($|W|=\rho\psi/2$). This idea can be traced back to the classical
investigations of \citet{Pl11} and \citet{Ed16}. It has received
additional impetus from the recent $N$-body simulations of
\citet{SI05}.

\subsection{Poisson's equation}

While we have established that any spherical system described by a
DF of the form of Eq.~(\ref{eq:ghv}) is
hypervirial, we still have to find the corresponding density and
potential by solving Poisson's equation (here $G=1$)
\begin{equation}
\frac{1}{r^2}\frac{\mathrm{d}}{\mathrm{d}r}
\left(r^2\frac{\mathrm{d}\psi}{\mathrm{d}r}\right)=
-4\pi\sum_iD_ir^{p_i-2}\psi^{2p_i+1}.
\label{eq:gle}
\end{equation}
%
%
The order of Eq.~(\ref{eq:gle}) can be reduced as follows
\citep[c.f.,][]{Ch39}. First, let us consider the substitution
$\psi=r^{-1/2}\varphi=\varphi\exp(-t/2)$ and $t=\ln r$. Then, the left
hand side of Eq.~(\ref{eq:gle}) transforms
\begin{equation}
\frac{1}{r^2}\frac{\mathrm{d}}{\mathrm{d}r}
\left(r^2\frac{\mathrm{d}\psi}{\mathrm{d}r}\right)
=r^{-5/2}\left(\frac{\mathrm{d}^2\varphi}{\mathrm{d}t^2}
-\frac{\varphi}{4}\right),
\end{equation}
while the right hand side can be rewritten using
\begin{equation}
\sum_iD_ir^{p_i-2}\psi^{2p_i+1}
=\sum_iD_ir^{-5/2}\varphi^{2p_i+1}.
\end{equation}
Hence, Eq.~(\ref{eq:gle}) reduces to
\begin{equation}
\frac{\mathrm{d}^2\varphi}{\mathrm{d}t^2}=
\frac{\varphi}{4}-4\pi\sum_iD_i\varphi^{2p_i+1},
\label{eq:redpoisson}
\end{equation}
which does not involve the independent variable explicitly, and so its
order can be reduced by standard techniques \citep[e.g.,][]{In44} to
give
\begin{equation}
\left(\frac{\mathrm{d}\varphi}{\mathrm{d}t}\right)^2
=A+\frac{\varphi^2}{4}\left(1-\sum_i4B_i\varphi^{2p_i}\right),
\label{eq:modpoisson}
\end{equation}
where $A$ is a constant of integration and
\begin{equation}
B_i=\frac{4\pi D_i}{p_i+1}.
\label{eq:con2}
\end{equation}
Using the boundary condition at infinity ($\varphi=0$ and
$\mathrm{d}\varphi/\mathrm{d}t=0$), we find that $A=0$.\footnote{The
solution with $A\ne0$ is unphysical because
Eq.~(\ref{eq:modpoisson}) implies that $(\mathrm{d}\ln\psi/\mathrm{d}\ln
r)$ diverges as $r\psi^2\rightarrow0$ unless $A=0$. Nevertheless, we
note that it is possible to find explicit solutions with $A\ne0$ if
the sum contains a single term with $p=1/2$, 1, or 2
(See Appendix~\ref{appen:gLE}). The resulting
solutions, expressible in terms of Jacobi or Weierstrass elliptic
functions, however, exhibit oscillatory behaviour along the real axis,
which we suspect to be a general property of the differential
equation.}  Then, after introducing a further transformation of the
variable, $\varphi^{-p}=\vartheta$, Eq.~(\ref{eq:modpoisson})
reduces to
\begin{equation}
\left(-\frac{2}{p}\frac{\mathrm{d}\vartheta}{\mathrm{d}t}\right)^2
=\vartheta^2-\sum_i4B_i\vartheta^{2(1-p_i/p)}.
\label{eq:deq}
\end{equation}
Here, we note that the differential equation $(y')^2=f(y)$ where
$f(y)$ is a polynomial of $y$, the degree of which is at most two, can
be solved through elementary functions. However, the right-hand side
of Eq.~(\ref{eq:deq}) can be a quadratic polynomial of
$\vartheta$ if $p=p_i$ or $p=2p_i$ for all distinct $p_i$'s. For the
simplest case when the sum contains only a single term, it is
straightforward to show that either choice of $p$ leads to the
solution that reduces to the models of \citet{EA05} with the
integration constant being the scalelength. The only other possibility
is that the sum contains two terms with $p=p_1=2p_2$, which is
investigated in the following section.

\subsection{Solutions}

For this case, the integration results in
\begin{equation}
\vartheta=2k\cosh\left[\frac{p}{2}(t\!-\!t_0)\right]+2B_2
=k\left[\left(\frac{r}{r_0}\right)^{p/2}
\!\!+\left(\frac{r}{r_0}\right)^{-p/2}\right]+2B_2,
\end{equation}
where $t_0=\ln r_0$ is the integration constant, and
$k=(B_1+B_2^2)^{1/2}$.
By reinstating $\vartheta=\varphi^{-p}=r^{-p/2}\psi^{-p}$, we obtain
\begin{equation}
\frac{1}{\psi^p}
=kr_0^{p/2}\left[1+2c\left(\frac{r}{r_0}\right)^{p/2}
\!\!+\left(\frac{r}{r_0}\right)^p\right],
\end{equation}
where $c=B_2/k$. With the normalization $M_\infty=1$, we find that
$k=r_0^{p/2}$. So, we arrive at a two-parameter -- $c$ and $p$ --
potential of the form of (incorporating the dimensional constants $G$,
$M$ and $r_0$)
\begin{equation}
\psi=\frac{GM}{\left(r_0^p+2cr_0^{p/2}r^{p/2}+r^p\right)^{1/p}},
\label{eq:nhv}
\end{equation}
whose DF is given by
\begin{equation}
f(E,L)=C_1L^{p-2}E^{3p/2+1/2}+C_2L^{p/2-2}E^{3p/4+1/2}.
\label{eq:ghvs}
\end{equation}
Here, the constants $C_i$ can be found from Eqs.~(\ref{eq:con1})
and (\ref{eq:con2}) with $B_1=1-c^2$, $B_2=c$, $p_1=p$ and $p_2=p/2$
(henceforth $G=M=r_0=1$), that is,
\begin{eqnarray*}
C_1&=&\frac{1-c^2}{2^{p/2+1}(2\pi)^{5/2}}
\frac{\Gamma(2p+3)}{\Gamma(p/2)\Gamma(3p/2+3/2)}\\
C_2&=&\frac{c}{2^{p/4+1}(2\pi)^{5/2}}
\frac{\Gamma(p+3)}{\Gamma(p/4)\Gamma(3p/4+3/2)}.
\end{eqnarray*}
In order for the DF to be
non-negative everywhere, we must have $p>0$ and $0\le c\le1$. In
particular, if $c=0$ or $c=1$, the models reduce to those of
\citet{EA05}. We note that the model with $(p,c)=(a,1)$ is identical
to the one with $(p,c)=(a/2,0)$.

\begin{figure}
\includegraphics[width=\hsize]{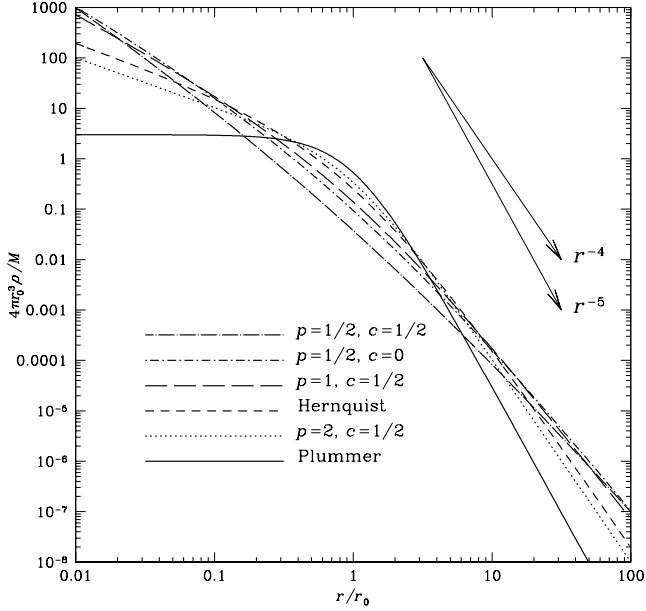}
\caption{\label{fig:genhyp} The density profile of the generalized
hypervirial models. The Plummer and Hernquist models are included in
the generalized hypervirial family as the cases $(p,c)=(2,0)$ [or
$(4,1)$] and $(p,c)=(1,0)$ [or $(2,1)$] respectively. Note that the
model with $(p,c)=(1/2,0)$ is the same as $(p,c)=(1,1)$.}
\end{figure}
\begin{figure}
\includegraphics[width=\hsize]{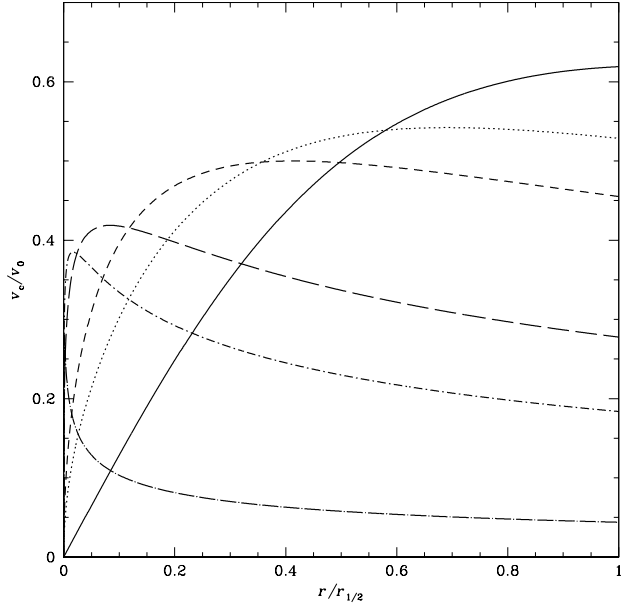}
\caption{\label{fig:genhypv} The circular velocity curves of the generalized
hypervirial models as a function of $r/r_{1/2}$, where $r_{1/2}$ is
the half-mass radius. Here, $v_0^2=GM/r_0$.
The line types are as in Fig.~1.}
\end{figure}

\section{The Simple halo models}
\label{sec:shm}

Thusfar, we have obtained the gravitational potential (Eq.~\ref{eq:nhv})
corresponding to the simple DF (Eq.~\ref{eq:ghvs}). Various limits are
already well-known. For example, when $(p,c)=(1,0)$ or $(2,1)$, this
is the \citet{He90} potential generated by the DF
first found by \citet{BD02}. When $(p,c)=(2,0)$ or $(4,1)$, this is the
isotropic \citet{Pl11} model. Bearing in mind the
property in Eq.~(\ref{eq:hyper}), we refer to this family as {\it the
generalized hypervirial models}.

The density generated by the potential of Eq.~(\ref{eq:nhv}) is
\begin{equation}
\rho=
\frac{(1\!-\!c^2)(p\!+\!1)(4\pi)^{-1}}{r^{2-p}(1\!+\!2cr^{p/2}\!+\!r^p)^{2+1/p}}
\!+\!\frac{c(p\!+\!2)(8\pi)^{-1}}{r^{2-p/2}(1\!+\!2cr^{p/2}\!+\!r^p)^{1+1/p}},
\label{eq:nhvd}
\end{equation}
which can be found from Poisson's equation. Fig.~\ref{fig:genhyp} shows
some typical density profiles.
Provided that $c\ne0$, the second term in Eq.~(\ref{eq:nhvd}) is
dominant when $r\rightarrow0$ [$\rho\sim r^{-(2-p/2)}$] and
$r\rightarrow\infty$ [$\rho\sim r^{-(3+p/2)}$]. If $p\le2$, the
density is monotonically decreasing outwards-radially regardless of
the value of $c$. If $2<p<4$, there may be a region of increasing
density depending on the value of $c$, although the model still
exhibits a cuspy centre. The $p=4$ model is cored whereas there is a
hole at the centre if $p>4$.

Notice that the density profile -- although composed of entirely
elementary functions -- is a bit more complicated than either the
potential or the DF. We argue that this is the right way around as most
applications will use the potential (for example, for integrating the
orbits in numerical simulations) or the DF (for
example, for calculating the flux of dark matter particles on a
detector). It is much more useful to have models with simple
potentials and DFs than those with simple density profile.

Evidence from N-body simulations suggests that the density
profile of the dark halo follows a simple functional form. One of the
most commonly cited examples is that of \citet*[henceforce
NFW]{NFW95,NFW96}, which is basically a double power-law characterized
by $r^{-1}$ cusp at the centre and $r^{-3}$ fall-off at large
radii. Since every member of the generalized hypervirial models has a
finite mass, none of them can reproduce the $r^{-3}$ density fall-off
-- which implies an infinite mass -- in the outer region. Regarding
the behaviour in the inner region, however, many members of the family
indeed exhibit a $r^{-1}$-like cusp, including the well-known example
of the \citet{He90} model. In fact, the additional freedom afforded by
the parameter $c$ admits more flexibility in the behaviour around the
scalelength. For example, we find that the model with $(p,c)=(2,3/4)$
provides a better fit to the NFW profile within a scalelength than
the \citet{He90} model. Actually, if we allow a slight deviation of
the cusp slope, there exists a trade-off between varying $p$ and $c$
which produces very similar behaviour of the density profiles in the
inner parts.

\begin{figure*}
\centering
\includegraphics[width=8cm]{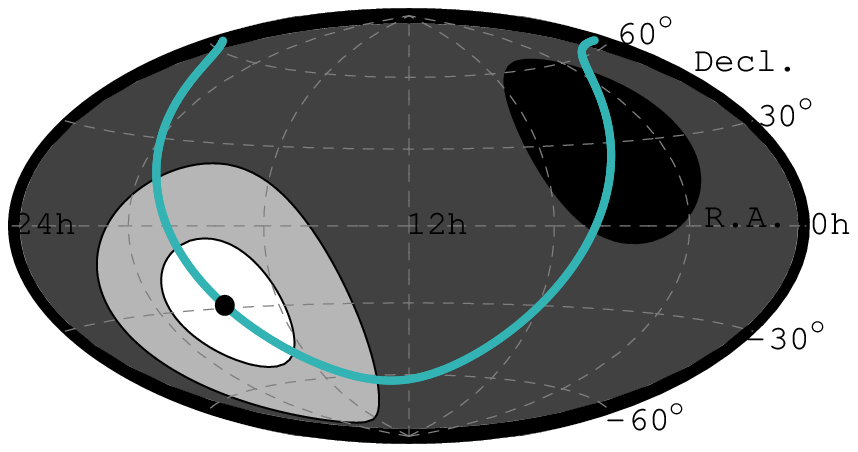}
\includegraphics[width=8cm]{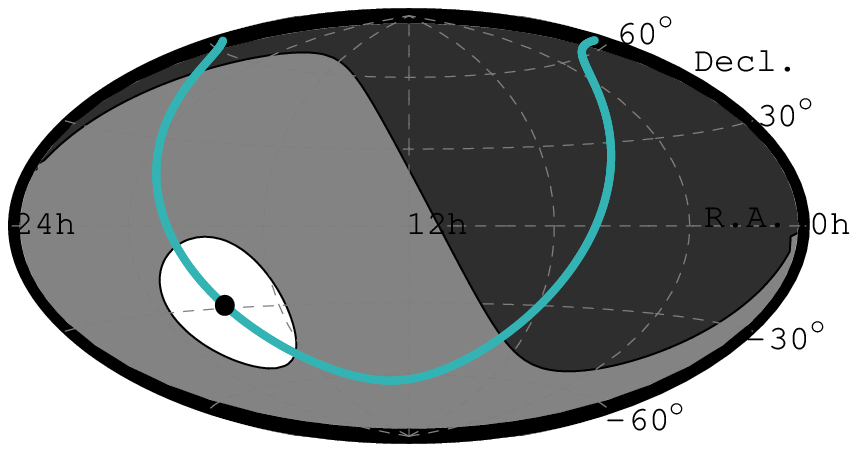}
\caption{\label{fig:hammer} Contour plots of the (projected) number
density of $\gamma$-ray sources in equatorial coordinates. The left
hand panel refers to a model with $\rho \propto r^{-1}$ at the centre
$(p,c)=(2,1/2)$, while the right-hand panel to a model with
$\rho\propto r^{-7/4}$ at the centre $(p,c)=(1/2,1/2)$. The grey-scale
gives the relative number density, with white representing the highest
values and black the lowest. The number density changes by a factor of
10 on moving from one contour to the next. The Galactic centre and the
Galactic plane are marked.}
\end{figure*}

The circular speed and the cumulative mass can be found as
\begin{equation}
v_\mathrm{c}^2=-r\frac{\mathrm{d}\psi}{\mathrm{d}r}
=\frac{cr^{p/2}+r^p}{(1+2cr^{p/2}+r^p)^{1/p+1}},
\end{equation}
\begin{equation}
M_r=rv_\mathrm{c}^2
=\frac{1+cr^{-p/2}}{(1+2cr^{-p/2}+r^{-p})^{1/p+1}}.
\end{equation}
Here, the total mass is finite and therefore the circular speed falls
off as Keplerian at large radii ($v_\mathrm{c}\sim
r^{-1/2}$). Fig.~\ref{fig:genhypv} shows plots of the circular velocity
as a function of $r/r_{1/2}$, where $r_{1/2}$ is the half-mass radius,
(i.e., $M_{r_{1/2}}=1/2$).
The models with inner density slopes $\rho\propto r^{-1}$ or $\rho
\propto r^{-3/2}$ have rotation curves similar to that of the NFW profile
or \citet{Mo98} models, and so are flattish over a wide range of radii.

The velocity dispersions are
\begin{equation}
\langle v_r^2\rangle =
\frac{1+c^2+cg}{2(1+c^2+cg)+p(2+cg)}\,\psi,
\label{eq:vrdis}
\end{equation}
\begin{equation}
\langle v_\mathrm{T}^2\rangle=\frac{p}{2}
\frac{2+cg}{2(1+c^2+cg)+p(2+cg)}\,\psi,
\label{eq:vthetadis}
\end{equation}
where $g=r^{p/2}+r^{-p/2}$ and therefore the anisotropy parameter
varies according to
\begin{equation}
\beta=1-\frac{p}{4}\,\left(\frac{2+cg}{1+c^2+cg}\right).
\end{equation}
From its construction, the hypervirial relation is automatically
satisfied. Because the potential is everywhere finite, the hypervirial
relation also implies that the velocity dispersions are everywhere
finite.
In particular, assuming $c\ne0$, the central velocity dispersions are
$\langle v_r^2\rangle=1/(p+2)$ and $\langle
v_\mathrm{T}^2\rangle=p/[2(p+2)]$. In addition,
\begin{equation}
\frac{d\beta}{dr}=
\frac{p^2c(1-c^2)r^{p/2-1}(r^p-1)}{8(c+r^{p/2})^2(1+cr^{p/2})^2},
\end{equation}
and therefore, provided that $0<c<1$, the anisotropy parameter $\beta$
decreases from $\beta=1-p/4$ at $r=0$ to $\beta=1-p/[2(1+c)]$ at $r=1$
and then increases back to $\beta\rightarrow1-p/4$ as
$r\rightarrow\infty$. In other words, in this model, the velocity
dispersions are more radially anisotropic near the centre and the
outskirts whereas they are relatively less radially anisotropic around
the region of the radial scalelength. However, we also note that the
model as a whole always possesses a more radially biased velocity
dispersion than the isotropic model unless $p>2(1+c)$.

It has been suggested that dark matter haloes achieve an almost
isotropic state near the centre and become more and more radially
anisotropic in the outer parts, at least according to cosmological
N-body simulations \citep{HM05}. Our DFs are radially anisotropic and
therefore better suited to modelling the outer parts and envelopes of
dark haloes. They are unsuitable for the class of problems in which
central anisotropy plays a critical role.

\begin{figure*}
\centering
\includegraphics[width=12cm]{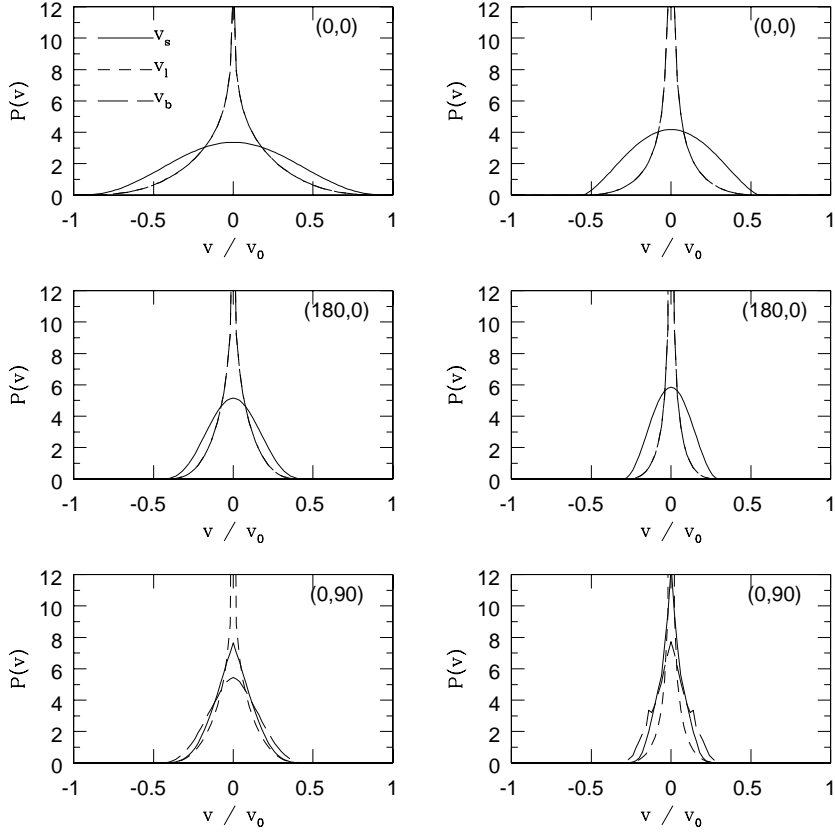}
\caption{\label{fig:proper} The distribution of line of sight
velocities and proper motions of dark matter particles towards the
Galactic Centre (upper panels), the anti-Centre (middle panels) and the
Galactic North Pole (lower panels). The left-hand panels refer to the
model with $(p,c)=(2,1/2)$, while the right-hand panels refer to the
model with $(p,c)=(1/2,1/2)$. The velocity is given in units of $v_0=
(GM/r_0)^{1/2}$.}
\end{figure*}

\section{An application: the distribution of $\gamma$-ray sources}
\label{sec:app}

\citet*{DMS05} have presented evidence from numerical
simulations that dark matter may be clumped into mini-haloes of Earth
mass and larger. They estimate that $\sim$50\% of the total
mass of the dark matter halo is bound to dark matter
substructures. These objects may be detectable by virtue of the
$\gamma$-rays from neutralino annihilation in the very centres of the
clumps. \citet{DMS05} also point out that the nearest
mini-haloes will be amongst the very brightest sources from neutralino
annihilation and may be found either with the forthcoming \textsl{GLAST}
satellite or next-generation atmospheric \v{C}erenkov telescopes as high
proper motion, discrete $\gamma$-ray sources.

If this idea is correct, then there are some immediate consequences.
First, because of the offset of the Sun's location from the centre of
the dark halo, the distribution of such $\gamma$-ray sources is
anisotropic and the magnitude of the anisotropy is an indicator of the
cusp slope at the Galactic Centre. This effect is already well-known
in studies of halo origin of the ultra-high energy cosmic rays
\citep*[e.g.,][]{EFS02}. Let us use Galactic coordinates
$(s,\ell,b)$, where $s=|\vec{s}|$ is heliocentric distance, and
$(\ell,b)$ are Galactic longitude and latitude. Then, the relative
number density of $\gamma$-ray sources is
\begin{equation}
F(\ell,b)\propto\int\!s^2\,\mathrm{d}s\,w(s)\,
\rho[\vec{R}_{\sun}+\vec{s}_{\ell,b}]
\label{eq:fluxint}
\end{equation}
where $\vec{R}_{\sun}$ is the Galactocentric solar position and $w(s)$ is
the selection function. For bright sources, the selection function is
proportional to the relative luminosity and so $w(s)\propto s^{-2}$.
The density profile is truncated at a radius at which the dark matter
annihilation rate matches the collapse timescale of the cusp. With
this assumption, a tiny constant density core is created
\citep[e.g.,][]{Ty02} and convergence of the integral~(\ref{eq:fluxint})
is guaranteed. The overall normalization of the integral depends
on the fraction of dark matter bound in mini-haloes, as opposed to
smoothly distributed dark matter. The anisotropy effect is clearly
illustrated in Fig.~\ref{fig:hammer}, which shows Hamer-Aitoff
projections in equatorial coordinates for two of the halo models. The
first is a model with $\rho\propto r^{-1}$ at the centre $(p,c)=
(2,1/2)$, and the second is a model with $\rho\propto r^{-7/4}$ at the
centre $(p,c)=(1/2,1/2)$. The more highly cusped the model, the greater
the anisotropy. If there are enough detections, the magnitude of the
anisotropy can be quantified by harmonic analysis \citep[see e.g.,][]{EFS02}.

Second, if the high proper motion sources can be identified, this will
provide the first direct evidence on the velocity distribution of the
dark matter. It is therefore useful to have predictions of the velocity
distributions. The proper motion and radial velocity
distribution of dark matter clumps can be calculated easily for
our models, because the DFs are simple. The proper motion and radial
velocity distributions in direction
$\mathrm{d}^2\Omega=\cos b\,\mathrm{d}\ell\,\mathrm{d}b$ is
\begin{equation}
\frac{\mathrm{d}^2N_\mathrm{obs}}{\mathrm{d}v_\ell\mathrm{d}v_b}
(v_\ell,v_b,\ell,b)=
\mathrm{d}^2\!\Omega\iint\!\mathrm{d}v_s\,\mathrm{d}s\
f(E,L)\,s^2\,w(s),
\end{equation}
\begin{equation}
\frac{\mathrm{d}N_\mathrm{obs}}{\mathrm{d}v_s}(v_s,\ell,b)=
\mathrm{d}^2\!\Omega
\iiint\!\mathrm{d}v_\ell\,\mathrm{d}v_b\,\mathrm{d}s\
f(E,L)\,s^2\,w(s).
\end{equation}
Here, the velocity of the dark matter particles has been resolved into
components $(v_s,v_\ell,v_b)$ based on Galactic coordinates while
$f(E,L)$ is the DF of the dark matter. For models
with DFs~(\ref{eq:ghv}), the above triple integrals are easily
computed via Gaussian quadrature. In fact, the second moments of
these distributions are entirely analytic, as sketched out in
Appendix~\ref{appen}. Fig.~\ref{fig:proper} shows the proper motion
and line of sight velocity distributions for the dark matter clumps in
three directions. When looking towards the Galactic Centre or
anti-Centre in spherical models, then the distributions of the two
tangential velocity components $v_\ell$ and $v_b$ are the same, though
this is not necessarily the case in other directions. All three
velocity distributions show very significant deviations from
Gaussianity. The sources with highest tangential velocity occur
towards the the Galactic Centre direction and it is therefore in this
direction that \textsl{GLAST} or \textsl{VERITAS} may find it easiest
to detect them unambiguously. 

Although these facts have been established in the context of our
simple family of models, we argue that they are likely to be
generic. The anisotropy effect in source positions is a consequence of
the Sun's offset from the centre. The fastest moving sources are
likely to be found in the direction in which the gravitational
potential well is deepest.

\section{Conclusions}

We have presented a simple family of halo models useful for the study
of dark matter haloes. They have a simple potential and distribution
function (DF), being composed of two terms of the form of
$L^{p-2}E^{(3p+1)/2}$ -- where $L$ is the total angular momentum, $E$
the binding energy and $p$ is some constant. As an application, we
have presented the properties of discrete $\gamma$-ray sources arising
from neutralino annihilation in dark matter mini-haloes. This
suggestion for the composition of the dark matter is derived the
numerical simulations of \citet{DMS05}. If true, then nearby, high
proper motion discrete $\gamma$-rays sources may be detectable by
forthcoming missions such as \textsl{GLAST}. We have shown that there
is an anisotropy in the positions of such $\gamma$-ray sources because
of the offset of the Sun from the Galactic Centre. We have also
provided distributions of proper motions and line of sight velocities
of such sources. The best direction in which to look is towards the
Galactic Centre $(\ell,b)=(0,0)$. There are more sources in this
direction and they have the highest proper motion.

This paper also continues the study of hyperviriality begun by
\citet{Ed16} and further developed by \citet{EA05}. In spherical
symmetry, any model which has a DF that is a sum of
terms like $L^{p-2}E^{(3p+1)/2}$ is hypervirial. This leads to the
derivation of a general differential equation that any spherical
hypervirial system must satisfy. We have solved it to find the most
general models, for which the potential can be written down in terms
of elementary functions.

We briefly note that, in the case of axisymmetry, any model which has
a DF that is a sum of terms like
$L_z^{p-2}E^{(3p+1)/2}$ -- where $L_z$ is the angular momentum
component about the axis of symmetry -- is also hypervirial
(see Appendix~\ref{appen:axis} for more details). It is
possible to derive the equation for the corresponding potential and
demonstrate that at least two analytical solutions exist,
corresponding to the flattened Plummer model studied by
\citet{Ly62} and a particular case among the prolate galaxy models of
\citet{La81}. In fact, only the Plummer model -- the sole isotropic,
hypervirial model -- can be generalized to give an axisymmetric
hypervirial system. It is an open question whether any further
hypervirial generalizations of the Plummer model can be found.

\onecolumn
\begin{appendix}

\section{Second moments of the velocity distributions}
\label{appen}

It is straightforward to compute the six independent components of the
velocity dispersion tensor in a Galactic coordinate system. These
formulae do not appear to have been given before, and so we quote them
here:

\begin{eqnarray}
\langle v_s^2\rangle&=&\langle v_x^2\rangle \cos^2\!b\,\cos^2\!\ell
                     + \langle v_y^2\rangle \cos^2\!b\,\sin^2\!\ell
                     + \langle v_z^2\rangle \sin^2\!b
                     + \langle v_xv_y\rangle\cos^2\!b\,\sin2\ell 
                     - \langle v_yv_z\rangle\sin2b\,   \sin\ell
                     - \langle v_xv_z\rangle\sin2b\,   \cos\ell,
\\%
\langle v_b^2\rangle&=&\langle v_x^2\rangle \sin^2\!b\,\cos^2\!\ell
                     + \langle v_y^2\rangle \sin^2\!b\,\sin^2\!\ell
                     + \langle v_z^2\rangle \cos^2\!b
                     + \langle v_xv_y\rangle\sin^2\!b\,\sin2\ell 
                     + \langle v_yv_z\rangle\sin2b\,   \cos\ell
                     + \langle v_xv_z\rangle\sin2b\,   \sin\ell,
\\%
\langle v_\ell^2\rangle&=&\langle v_x^2 \rangle\sin^2\!\ell
                        + \langle v_y^2 \rangle\cos^2\!\ell
                        - \langle v_xv_y\rangle\sin2\ell,
\end{eqnarray}
\begin{eqnarray}
\langle v_sv_b\rangle
 &=&-\langle v_x^2\rangle \cos b\,\sin b\,\cos^2\!\ell
    -\langle v_y^2\rangle \cos b\,\sin b\,\sin^2\!\ell
    +\langle v_z^2\rangle \cos b\,\sin b
    -\langle v_xv_y\rangle\sin2b\,\cos\ell\,\sin\ell\cr
 & &-\langle v_xv_z\rangle\cos2b\,\cos\ell 
    -\langle v_yv_z\rangle\cos2b\,\sin\ell,
\\%
\langle v_\ell v_s\rangle
 &=&\langle v_y^2-v_x^2\rangle\cos b\,\sin\ell\,\cos\ell
  + \langle v_xv_y\rangle     \cos b\,\cos2\ell
  + \langle v_xv_z\rangle     \sin b\,\sin\ell
  - \langle v_yv_z\rangle     \sin b\,\cos\ell,
\\%
\langle v_\ell v_b\rangle
 &=&\langle v_x^2-v_y^2\rangle\sin b\,\sin\ell\,\cos\ell
  - \langle v_xv_y\rangle     \cos b\,\cos2\ell
  + \langle v_xv_z\rangle     \cos b\,\sin\ell
  - \langle v_yv_z\rangle     \cos b\,\cos\ell.
\end{eqnarray}
This gives the observables in terms of the second moments referred to
Cartesian coordinates. The latter can be related to the second moments
in spherical polars (in which the tensor diagonalizes) via:
\begin{equation}
\langle v_x^2 \rangle
=\frac{x^2}{r^2}     \langle v_r^2      \rangle
+\frac{y^2+z^2}{r^2} \langle v_\theta^2 \rangle,
\qquad%
\langle v_y^2 \rangle 
=\frac{y^2}{r^2}     \langle v_r^2      \rangle
+\frac{x^2+z^2}{r^2} \langle v_\theta^2 \rangle,
\qquad%
\langle v_z^2 \rangle
=\frac{z^2}{r^2}     \langle v_r^2      \rangle
+\frac{x^2+y^2}{r^2} \langle v_\theta^2 \rangle,
\end{equation}
\begin{equation}
\langle v_xv_y\rangle
=\frac{xy}{r^2} 
\left(\langle v_r^2\rangle-\langle v_\theta^2\rangle\right),
\qquad%
\langle v_xv_z \rangle
=\frac{xz}{r^2}
\left(\langle v_r^2\rangle-\langle v_\theta^2\rangle\right),
\qquad%
\langle v_yv_z \rangle
=\frac{yz}{r^2}
\left(\langle v_r^2\rangle-\langle v_\theta^2\rangle\right).
\end{equation}
Here, $\langle v_\theta^2\rangle=\langle v_\phi^2\rangle=\langle
v_\mathrm{T}^2\rangle/2$. Hence, by substituting in the known velocity
dispersions in spherical polars (Eqs.~\ref{eq:vrdis} and
\ref{eq:vthetadis}), we obtain the second moments of the observable
distributions (the line of sight velocities and the proper motions).

\input aa3647-05-appB

\input aa3647-05-appC

\end{appendix}

\end{document}

%% file: aa3647-05-appB
\section{Axisymmetric hypervirial potentials}
\label{appen:axis}

Upon the discovery of the new hypervirial potential, one may ask the
question: does there exist a simple extension of the spherical
hypervirial potentials into axisymmetry? A possible starting point of
the investigation is the examination of axisymmetric DFs
that are obtained by replacing the $L$-dependence of a
spherically symmetric DF with an $L_z$-dependence.
Let us suppose that the DF of an axisymmetric
system is given by
\begin{equation}
f(E,L_z)=C|L_z|^{2n}E^{m-3/2},
\label{eq:asdf}
\end{equation}
where $L_z=Rv_\phi$ is the angular momentum along the symmetry axis,
and $n>-1/2$ and $m>1/2$. Then, a simple integration leads to the
expressions for the corresponding density \citep[c.f.,][]
{Fr52,De86,Ev94}
\begin{equation}
\rho=2^{n+3/2}\pi C
\frac{\Gamma(n+1/2)\Gamma(m-1/2)}{\Gamma(n+m+1)}R^{2n}\psi^{n+m},
\label{eq:aden}
\end{equation}
and the second-order velocity moments
\begin{eqnarray}\lefteqn{
\rho\langle v_R^2\rangle=2^{n+3/2}\pi C
\frac{\Gamma(n+1/2)\Gamma(m-1/2)}{\Gamma(n+m+2)}R^{2n}\psi^{n+m+1}
=\frac{\rho\psi}{n+m+1}\,;}
\nonumber\\\lefteqn{
\rho\langle v_\phi^2\rangle=2^{n+5/2}\pi C
\frac{\Gamma(n+3/2)\Gamma(m-1/2)}{\Gamma(n+m+2)}R^{2n}\psi^{n+m+1}
=(2n+1)\rho\langle v_R^2\rangle.}
\label{eq:asvel}
\end{eqnarray}
Note that $\rho\langle v_R^2\rangle=\rho\langle v_z^2\rangle=
\rho\langle v_r^2\rangle=\rho\langle v_\theta^2\rangle$ since the
DF is symmetric with respect to $v_R\leftrightarrow
v_z$ exchange. Finally, we find that, for the system described by the
DF of Eq.~(\ref{eq:asdf}),
\begin{equation}
\rho\langle v^2\rangle=\rho\left(\langle v_R^2\rangle
+\langle v_z^2\rangle+\langle v_\phi^2\rangle\right)
=(2n+3)\rho\langle v_R^2\rangle=\frac{2n\!+\!3}{n\!+\!m\!+\!1}\ \rho\psi,
\end{equation}
and thus that the steady-state virial theorem holds without the
boundary terms only if $m=3n+5$, for which the system becomes
hypervirial. Here, the density of the system (Eq.~\ref{eq:aden}) is
axisymmetric, but there is no net rotation ($\langle
v_\phi\rangle=0$), as the DF~(\ref{eq:asdf}) is
symmetric with respect to $L_z\leftrightarrow -L_z$ exchange. To
rectify this, we can amend the DF~(\ref{eq:asdf}) by
%
\begin{equation}
f(E,L_z)=C\left[1+\xi(L_z)\right]|L_z|^{2n}E^{m-3/2}
\label{eq:odf}
\end{equation}
where $\xi(L_z)$ is an arbitrary odd function of $L_z$ [i.e,
$\xi(-L_z)=-\xi(L_z)$] that is bounded by $-1\le\xi(L_z)\le1$. Note
that, on the ground of statistical mechanics, \citet{De86,De87apj}
advocated the form of $\xi(L_z)$ being $\tanh(kL_z)$ where $k$ is a
parameter that depends on the total angular momentum along the
symmetry axis. Physically, this action of adding a certain odd
function to the even distribution function corresponds to switching
the direction of rotations for a certain fraction of stars of given
$|L_z|$ and $E$. It is straightforward to show that, for the DF of
Eq.~(\ref{eq:odf}), Eqs.~(\ref{eq:aden}) and
(\ref{eq:asvel}) are still true while
\begin{eqnarray*}
\rho\langle v_\phi\rangle=\int\!v_\phi f\,\mathrm d^3\!\vec{v}
&=&\frac{4\pi C}{2m-1}\,
\int_{-\sqrt{2\psi R^2}}^{\sqrt{2\psi R^2}}\!\frac{\mathrm dL_z}{R^2}\,L_z
\left[1+\xi(L_z)\right]\left(L_z^2\right)^n
\left(\psi-\frac{L_z^2}{2R^2}\right)^{m-1/2}\\
&=&\frac{2^{n+3}\pi C}{2m-1}\,R^{2n}\psi^{n+m+1/2}\int_0^1\!
\xi\left(\sqrt{2\psi R^2t}\right)\,t^n\left(1-t\right)^{m-1/2}\,\mathrm dt\\
&\le&2^{n+2}\pi C
\frac{\Gamma(n+1)\Gamma(m-1/2)}{\Gamma(n+m+3/2)}R^{2n}\psi^{n+m+1/2}
=2^{1/2}
\frac{\Gamma(n+1)\Gamma(n+m+1)}{\Gamma(n+1/2)\Gamma(n+m+3/2)}\rho\psi^{1/2}.
\end{eqnarray*}
In other words, one can choose $\xi(L_z)$ to meet the constraint on
the net rotation $\langle v_\phi\rangle$. Here, the inequality becomes
the equality if $\xi(L_z)=1$ for $L_z>0$, which corresponds to the
situation when every star rotates in the same direction. For $n>-1/2$
and $m=3n+5$, it is straightforward to established that $\langle
v_\phi\rangle^2<\psi/2$.

Systems given by the DF~(\ref{eq:asdf}) are
somewhat unrealistic since the density along the symmetry axis is
either zero ($n>0$) or infinite ($n<0$) except the $n=0$ case that is
just the isotropic (and spherically symmetric) Plummer model. However,
analogous to the procedure in Sect.~\ref{sec:df}, we may obtain
more realistic systems by summing DFs of the form
of Eq.~(\ref{eq:asdf}), one of which corresponds to that of the
isotropic Plummer model, that is,
\begin{equation}
f(E,L_z)=C_0E^{7/2}+\sum_iC_i|L_z|^{2n_i}E^{3n_i+7/2}
+f_\mathrm{odd}(E,L_z)
\end{equation}
where each $n_i$ should be positive in order to avoid the divergent
behaviour of the density along the symmetry axis. Here,
$f_\mathrm{odd}(E,L_z)$ is an arbitrary function that satisfies the
condition that $f_\mathrm{odd}(E,-L_z)=-f_\mathrm{odd}(E,L_z)$ and
chosen so that the $f(E,L_z)$ is non-negative for all accessible $E$
and $L_z$. It is a simple exercise to find expressions for the density
\begin{equation}
\rho=D_0\psi^5+\sum_iD_iR^{2n_i}\psi^{4n_i+5}
\label{eq:asden}
%
\,;\qquad
%
D_0=\frac{7\pi^2}{2^{11/2}}C_0\,;\qquad D_i=2^{n_i+3/2}\pi C_i
\frac{\Gamma(n_i+1/2)\Gamma(3n_i+9/2)}{\Gamma(4n_i+6)},
\end{equation}
and the second-order velocity moments
\begin{equation}
\rho\langle v_R^2\rangle=
\frac{D_0}{6}\psi^6+\sum_i\frac{D_i}{2(2n_i+3)}\
R^{2n_i}\psi^{4n_i+6}\,;\qquad
\rho\langle v_\phi^2\rangle=
\frac{D_0}{6}\psi^6+\sum_i\frac{(2n_i+1)D_i}{2(2n_i+3)}\
R^{2n_i}\psi^{4n_i+6}.
\end{equation}
It is also easy to establish that the system is in fact hypervirial,
\begin{equation}
\rho\langle v^2\rangle=
\rho\left(2\langle v_R^2\rangle+\langle v_\phi^2\rangle\right)=
\frac{D_0}{2}\psi^6+\sum_i\frac{D_i}{2}\
R^{2n_i}\psi^{4n_i+6} =\rho\frac{\psi}{2}.
\end{equation}

However, finding the corresponding potential is actually a non-trivial
problem as it is tantamount to solving Poisson's equation, which now
becomes a second-order partial differential equation:
\begin{equation}
\frac{1}{r^2}\frac{\partial}{\partial r}
\left(r^2\frac{\partial\psi}{\partial r}\right)+
\frac{1}{r^2\sin\theta}\frac{\partial}{\partial r}
\left(\sin\theta\frac{\partial\psi}{\partial\theta}\right)= -4\pi
D_0\psi^5-4\pi\sum_iD_ir^{2n_i}\sin^{2n_i}\!\theta\,\psi^{4n_i+5}
\end{equation}
in the spherical polar coordinate system or
\begin{equation}
\frac{1}{R}\frac{\partial}{\partial R}
\left(R\frac{\partial\psi}{\partial R}\right)+
\frac{\partial^2\psi}{\partial z^2}= -4\pi
D_0\psi^5-4\pi\sum_iD_iR^{2n_i}\psi^{4n_i+5}
\end{equation}
in the cylindrical polar coordinate system. Even after factoring out
the scalefree contribution \citep[c.f,][]{To82},
\begin{equation}
\frac{\partial^2\varphi}{\partial t^2}+
\cosh^2\!\zeta\,\frac{\partial^2\varphi}{\partial\zeta^2}
-\frac{\varphi}{4}=-4\pi D_0\varphi^5
-4\pi\sum_iD_i\frac{\varphi^{4n_i+5}}{\cosh^{2n_i}\!\zeta}
\end{equation}
where $\varphi=r^{1/2}\psi$, $t=\ln r$, and $\cos\theta=\tanh\zeta$ or
\begin{equation}
R^2\left(\frac{\partial^2\phi}{\partial R^2}+
\frac{\partial^2\phi}{\partial z^2}\right)+\frac{\phi}{4}=
-4\pi D_0\phi^5-4\pi\sum_iD_i\phi^{4n_i+5}
\end{equation}
where $\phi=R^{1/2}\psi$, these are not obviously separable in either
coordinate system. Instead, here, we take a rather different approach.
Inspired by Eq.~(\ref{eq:nhv}), we first consider the
axisymmetric potential of the form of
\begin{equation}
\psi=\frac{1}{[1+c(\theta)r^{p/2}+r^p]^{1/p}}
\end{equation}
where $c(\theta)$ is some function of the latitude $\theta$. Then, the
corresponding density can be found to be ($G=1$)
\begin{equation}
\rho=-\frac{1}{4\pi}\nabla^2\psi=
(p+1)\left[1-\frac{c^2}{4}
-\left(\frac{1}{p}\frac{\mathrm{d}c}{\mathrm{d}\theta}\right)^2\right]
r^{p-2}\psi^{2p+1}
+\frac{1}{p}\left[\frac{1}{\sin\theta}\frac{\mathrm{d}}{\mathrm{d}\theta}
\left(\sin\theta\frac{\mathrm{d}c}{\mathrm{d}\theta}\right)
+\frac{p(p+2)}{4}c\right]
r^{p/2-2}\psi^{p+1}.
\end{equation}
Now, we find that one of the necessary condition for the system to be
consistent with the hypervirial DF is that
$c(\theta)$ is the solution of the inhomogeneous Legendre equation
\begin{equation}
\frac{1}{\sin\theta}\frac{\mathrm{d}}{\mathrm{d}\theta}
\left(\sin\theta\frac{\mathrm{d}c}{\mathrm{d}\theta}\right)
+\frac{p}{2}\left(\frac{p}{2}+1\right)c=pC\sin^{p/2-2}\!\theta
\end{equation}
where $C$ is a non-negative constant. A simple trial reveals that
$c_0=4Cp^{-1}\sin^{p/2}\!\theta$ is its particular solution,
and so the general solution is
\begin{equation}
c=AP_{p/2}(\cos\theta)+BQ_{p/2}(\cos\theta)+\frac{4C}{p}\sin^{p/2}\!\theta
\label{eq:gls}
\end{equation}
where $P_l(x)$ is the Legendre-P function (polynomial if $l$ is an
integer), $Q_l(x)$ is the Legendre-Q function, and $A$ and $B$ are the
integration constants. Finally, the system can be hypervirial only if
\begin{equation}
1-\frac{c^2}{4}
-\left(\frac{1}{p}\frac{\mathrm{d}c}{\mathrm{d}\theta}\right)^2
\propto\sin^{p-2}\!\theta
\end{equation}
where $c$ is given by Eq.~(\ref{eq:gls}). It is rather obvious
that we only need to examine the cases when $p/2$ is an integer and
$B=0$. After a further examination, we find that, only possible cases
are $p=4$
\begin{equation}
c(\theta)=3\cos^2\!\theta-1+C\sin^2\!\theta
=2+(C-3)\sin^2\!\theta\,;
\qquad
\psi=\frac{1}{[1+2r^2+r^4+(C-3)R^2]^{1/4}},
\label{eq:lake}
\end{equation}
and $p=2$
\begin{equation}
c(\theta)=A\cos\theta+2C\sin\theta\,;
\qquad
\psi=\frac{1}{[1+r^2+Az+2CR]^{1/2}}
\end{equation}
where $A^2\le4(1-C^2)$. The corresponding densities for both cases
reduce the form of Eq.~(\ref{eq:asden}). In fact, the potential
corresponding to both solutions are already known!

The first ($p=4$) is the flattened Plummer model of \citet{Ly62},
%
\begin{equation}
\psi=\frac{\psi_0a}{[(r^2+a^2)^2-2b^2R^2]^{1/4}}\,;\qquad
%
%
\rho
=\frac{\psi_0(3-2\epsilon)}{\pi Ga^2}\left(\frac{\psi}{\psi_0}\right)^5
+\frac{5\psi_0\epsilon(2-\epsilon)}{\pi Ga^2}
\left(\frac{R}{a}\right)^2\left(\frac{\psi}{\psi_0}\right)^9
\end{equation}
where $0\le\epsilon=b^2/a^2\le3/2$. Comparing this to
Eq.~(\ref{eq:asden}), we find that the DF
\begin{equation}
f_\mathrm{even}(E,L_z)= C_0E^{7/2}+C_1L_z^2E^{13/2}
\,;\qquad
C_0=\frac{2^{11/2}}{7\pi^3}\ \frac{3-2\epsilon}{Ga^2\psi_0^4}\,;\qquad
C_1=\frac{2^{25/2}\cdot3\cdot5}{11\cdot13\pi^3}\
\frac{\epsilon(2-\epsilon)}{Ga^4\psi_0^8}
\end{equation}
can build the potential-density pair of the flattened Plummer model.
The second-order velocity moments are
\begin{equation}
\rho\langle v_R^2\rangle=
\frac{\psi_0^2(3-2\epsilon)}{6\pi Ga^2}\
\left(\frac{\psi}{\psi_0}\right)^6+
\frac{\psi_0^2\epsilon(2-\epsilon)}{2\pi Ga^2}\
\left(\frac{R}{a}\right)^2\left(\frac{\psi}{\psi_0}\right)^{10}\,;
\qquad
\rho\langle v_\phi^2\rangle=
\frac{\psi_0^2(3-2\epsilon)}{6\pi Ga^2}\
\left(\frac{\psi}{\psi_0}\right)^6+
\frac{3\psi_0^2\epsilon(2-\epsilon)}{2\pi Ga^2}\
\left(\frac{R}{a}\right)^2\left(\frac{\psi}{\psi_0}\right)^{10},
\end{equation}
and the system is hypervirial, as already noted by \citet{Ly62}.

The second ($p=2$) is a particular case of the generalized Plummer
model devised by \citet{La81}, for which 
\begin{equation}
\psi=\frac{\psi_0a}{[r^2+a^2+2bR]^{1/2}}\,;\qquad
%
%
\rho
=\frac{3\psi_0(1-\epsilon^2)}{4\pi Ga^2}
\left(\frac{\psi}{\psi_0}\right)^5
+\frac{\psi_0\epsilon}{4\pi G a^2} 
\left(\frac{a}{R}\right)\left(\frac{\psi}{\psi_0}\right)^3
\end{equation}
where $0\le\epsilon=b/a\le1$. This reduces to Eq.~(\ref{eq:lake})
after an arbitrary translation along the symmetry axis ($R=0$). In
fact, this is singular on the entire $R=0$ axis and so best represents
prolate galaxies. The DF is
\begin{equation}
f_\mathrm{even}(E,L_z)= C_0E^{7/2}+C_1 \delta(L_z)E^2
\,;\qquad
C_0=\frac{2^{7/2}\cdot 3}{7\pi^3}\ \frac{1-\epsilon^2}{Ga^2\psi_0^4}\,;\qquad
C_1=\frac{3}{8\pi^2}\frac{\epsilon}{G a\psi_0^2}.
\end{equation}
Note that the Dirac-$\delta$ distribution is the limiting case of
Eq.~(\ref{eq:asdf}) as in
$\lim_{n\rightarrow(-1/2)}[|L_z|^{2n}/\Gamma(n+1/2)]=\delta(L_z)$. The
second-order velocity moments are
\begin{equation}
\rho\langle v_R^2\rangle
=\frac{\psi_0^2(1-\epsilon^2)}{8\pi Ga^2}
\left(\frac{\psi}{\psi_0}\right)^6
+\frac{\psi_0^2\epsilon}{16\pi G a^2}
\left(\frac{a}{R}\right)\left(\frac{\psi}{\psi_0}\right)^4\,;
\qquad
\rho\langle v_\phi^2\rangle
=\frac{\psi_0^2(1-\epsilon^2)}{8\pi Ga^2}
\left(\frac{\psi}{\psi_0}\right)^6,
\end{equation}
and so this model is hypervirial.

The existence of the flattened Plummer models poses the question as to
whether there are axisymmetric generalizations with simple
DFs for all the hypervirial models. We argue that
this is not the case because the Plummer model is the only isotropic
hypervirial model. Any DF of $f(E,L)$ of a
spherically symmetric system has degenerate velocity dispersions in
the tangential plane, while any DF $f(E,L_z)$ of a
axisymmetric system has the velocity dispersions within the meridional
plane degenerate. Clearly, only the isotropic case can have both the
tangential plane and meridional plane as planes of symmetry of the
local velocity ellipsoid. This, therefore, implies that, if the
flattening method, however it is chosen, is gradual in the sense that
it includes the spherically symmetric case as a particular case, the
corresponding DF should necessarily reduce to not
only a spherically symmetric but also an isotropic DF
of $f(E)$. In other words, only those spherically symmetric
potentials with a simple \emph{isotropic} DF can
have an axisymmetric family of its generalization with continuous
values of flattening parameters. This, of course, does not rule out
the possibility that with a very special values of flattening
parameters, the corresponding DF may be simple but
there would not be any continuous transitions to the spherically
symmetric case while maintaining the simplicity of the DF.

%% file: aa3647-05-appC
\section{General solutions of generalized Lane-Emden equation}
\label{appen:gLE}

The differential equation in Eq.~(\ref{eq:gle}) when the sum contains only
a single term is a particular case of a family of differential equations
investigated by \citet{GH00}, which they called the generalized
Lane-Emden equation. In fact, Eq.~(\ref{eq:gle}) corresponds to
the special case of parameter combinations noted by them \citep[see
eq.~11 of][]{GH00} that permits a rational transformation of variables
that leaves the differential equation form-invariant. The
one-parameter solution family \citep[eq.~18 of][]{GH00} is the
direct generalization of the Schuster-Emden integral \citep[see e.g.]
[]{Ho86} and corresponds to the solution for the case of $A=0$ in
Eq.~(\ref{eq:modpoisson}), which leads to the models of
\citet{EA05}.

If we consider a transformation, $\eta=\varphi^s$,
Eq.~(\ref{eq:modpoisson}) with a single term in the sum further
reduces to
\begin{equation}
\left(\frac{\mathrm d\eta}{\mathrm dt}\right)^2=
As^2\eta^{2-2/s}+\frac{s^2}{4}\eta^2-Bs^2\eta^{2+2p/s}.
\label{eq:a1}
\end{equation}
We note that the differential equation of the form of $(y')^2=f(y)$
where $f(y)$ is a cubic or quartic polynomial of $y$ can be solved
with standard elliptic functions. Here, for $AB\ne0$, the right-hand
side of Eq.~(\ref{eq:a1}) can become a cubic of quartic
polynomial of $\eta$ if $s=\pm1$ (with $p=1$ or $p=2$) or $s=\pm2$
(with $p=1/2$ or $p=1$). In other words, the two-parameter general
solutions of Eq.~(\ref{eq:gle}), or equivalently the solutions of
Eq.~(\ref{eq:modpoisson}) with $A\ne0$ can be written down in
closed form using elliptic functions if the sum contains a single term
with $p=$1/2, 1, or 2 \citep{GH00}.

\subsection{Solutions in terms of Weierstrass-P functions}

Let us be reminded that the Weierstrass-P function is the canonical
elliptic function (i.e., complex bi-periodic meromorphic function) with
second-order poles. It is usually defined in terms of the sum over the
lattice points in complex plane with two half-periods. However, for
our purpose, it is useful to consider the differential equation
\begin{equation}
\left[\frac{\mathrm d}{\mathrm dz}\wp(z;g_2,g_3)\right]^2
=4[\wp(z;g_2,g_3)]^3-g_2\wp(z;g_2,g_3)-g_3
\label{eq:wpdep}
\end{equation}
that is satisfied by $\wp(z;g_2,g_3)$. Here, $g_2$ and $g_3$ are
usually referred to as elliptic invariants. We further note that the
differential equation (\ref{eq:wpdep}) does not explicitly involve the
independent variable so that one integration constant is given by an
arbitrary translation, that is, if $\varphi(t)$ is its solution, then
$\varphi(t+c)$ where $c$ is an arbitrary (complex) constant is also
the solution. However, Eq.~(\ref{eq:wpdep}) is a first order
differential equation, and thus, $\wp(z+c;g_2,g_3)$ is indeed the
complete specification of its general solution. In addition, the above
differential equation also indicates that $\wp$-function is
homogeneous, i.e.,
\begin{equation}
\wp(\lambda z;g_2,g_3)=\lambda^{-2}\wp(z;\lambda^4g_2,\lambda^6g_3).
\end{equation}

The right-hand side of Eq.~(\ref{eq:a1}) becomes a cubic
polynomial if i) $B\ne0$, $s=2p$ ($p=1/2$ or 1) or ii) $A\ne0$, $s=-2$
($p=1$ or 2). For these cases, linear transformations $\eta=\lambda
(\zeta+\tau)$ with properly chosen $\lambda$ and $\tau$ can reduce
Eq.~(\ref{eq:a1}) to the form of Eq.~(\ref{eq:wpdep}).
After a bit of algebra, we find solutions of
Eq.~(\ref{eq:modpoisson}) \citep[see also][]{GH00};
\begin{equation}
\varphi=\frac{1}{B}\left[\frac{1}{2^2\cdot3}
-2^2\wp\left(t+c;\,\frac{1}{2^6\cdot3},\,
-\frac{AB^2}{2^4}-\frac{1}{2^9\cdot3^3}\right)\right]
\qquad\ \mbox{for }\ B\ne0,\ p=1/2,
\end{equation}
\begin{equation}
\varphi^2=\frac{1}{B}\left[\frac{1}{2^2\cdot3}
-\wp\left(t+c;\,2^2AB+\frac{1}{2^2\cdot3},\,
-\frac{AB}{3}-\frac{1}{2^3\cdot3^3}\right)\right]
\qquad\ \mbox{for }\ B\ne0,\ p=1,
\end{equation}
\begin{equation}
\varphi^{-2}=\frac{1}{A}\left[\wp\left(t+c;\,2^2AB+\frac{1}{2^2\cdot3},\,
-\frac{AB}{3}-\frac{1}{2^3\cdot3^3}\right)
-\frac{1}{2^2\cdot3}\right]
\qquad\ \mbox{for }\ A\ne0,\ p=1,
\end{equation}
\begin{equation}
\varphi^{-2}=\frac{1}{A}\left[\wp\left(t+c;\,\frac{1}{2^2\cdot3},\,
2^2A^2B-\frac{1}{2^3\cdot3^3}\right)
-\frac{1}{2^2\cdot3}\right]
\qquad\ \mbox{for }\ A\ne0,\ p=2.
\end{equation}

\subsection{Solutions for $p=1$ in terms of Jacobi elliptic functions}

The Jacobi elliptic functions are defined through the inverse function
of the elliptic integral of the first kind, i.e.,
\begin{equation}
\mbox{sn}(u,k)=\sin\phi\,;\qquad
\mbox{cn}(u,k)=\cos\phi\,;\qquad
\mbox{dn}(u,k)=\sqrt{1-k^2[\mbox{sn}(u,k)]^2}
\end{equation}
where
\begin{equation}
u=F(\phi,k)=\int_0^\phi\!\frac{\mathrm dt}{\sqrt{1-k^2\sin^2\!t}}
=\int_0^{\mbox{sn}(u,k)}\!\frac{\mathrm dv}{(1-v^2)^{1/2}(1-k^2v^2)^{1/2}}
\end{equation}
and $k$ is the elliptic modulus. Remaining elliptic functions are
defined as
\begin{equation}
\begin{array}{lll}
\mbox{ns}\,u\equiv\mbox{sn}\,u;&
\mbox{nc}\,u\equiv\mbox{cn}\,u;&
\mbox{nd}\,u\equiv\mbox{dn}\,u;\\
\mbox{sc}\,u\equiv\mbox{sn}\,u\ \mbox{nc}\,u;\qquad&
\mbox{cd}\,u\equiv\mbox{cn}\,u\ \mbox{nd}\,u;\qquad&
\mbox{ds}\,u\equiv\mbox{dn}\,u\ \mbox{ns}\,u;\qquad\\
\mbox{cs}\,u\equiv\mbox{ns}\,u\ \mbox{cn}\,u;&
\mbox{dc}\,u\equiv\mbox{nc}\,u\ \mbox{dn}\,u;&
\mbox{sd}\,u\equiv\mbox{nd}\,u\ \mbox{sn}\,u.
\end{array}
\end{equation}
Here, the common elliptic modulus $k$ is omitted following usual
practices. The derivative of $\mbox{sn}\,u$ is given by
\begin{equation}
\frac{\mathrm d}{\mathrm du}\,\mbox{sn}\,u=\mbox{cn}\,u\ \mbox{dn}\,u
\label{eq:jed}
\end{equation}
and the derivatives of remaining elliptic functions can be found using
their respective definitions and Eq.~(\ref{eq:jed}). Then, it is
straightforward to show that each Jacobi elliptic function is a particular
solution of second-order differential equations such that
\begin{equation}
\begin{array}{lll}
(\mbox{sn}\,u)^{\prime\prime}=-(1+k^2)\,\mbox{sn}\,u+2k^2\,\mbox{sn}^3u;&
(\mbox{cn}\,u)^{\prime\prime}=(2k^2-1)\,\mbox{cn}\,u-2k^2\,\mbox{cn}^3u;&
(\mbox{dn}\,u)^{\prime\prime}=(2-k^2)\,\mbox{dn}\,u-2\,\mbox{dn}^3u;\\
(\mbox{ns}\,u)^{\prime\prime}=-(1+k^2)\,\mbox{ns}\,u+2\,\mbox{ns}^3u;&
(\mbox{nc}\,u)^{\prime\prime}=(2k^2-1)\,\mbox{nc}\,u+2(1-k^2)\,\mbox{nc}^3u;&
(\mbox{nd}\,u)^{\prime\prime}=(2-k^2)\,\mbox{nd}\,u-2(1-k^2)\,\mbox{nd}^3u;\\
(\mbox{cd}\,u)^{\prime\prime}=-(1+k^2)\,\mbox{cd}\,u+2k^2\,\mbox{cd}^3u;&
(\mbox{sd}\,u)^{\prime\prime}=(2k^2-1)\,\mbox{sd}\,u-2k^2(1-k^2)\,\mbox{sd}^3u;&
(\mbox{sc}\,u)^{\prime\prime}=(2-k^2)\,\mbox{sc}\,u+2(1+k^2)\,\mbox{sc}^3u;\\
(\mbox{dc}\,u)^{\prime\prime}=-(1+k^2)\,\mbox{dc}\,u+2\,\mbox{dc}^3u;&
(\mbox{ds}\,u)^{\prime\prime}=(2k^2-1)\,\mbox{ds}\,u+2\,\mbox{ds}^3u;&
(\mbox{cs}\,u)^{\prime\prime}=(2-k^2)\,\mbox{cs}\,u+2\,\mbox{cs}^3u
\end{array}
\label{eq:deqel}
\end{equation}
where the primed symbols indicate the differentiation with respect to
the argument $u$ and the simplified notations for power (e.g.,
$\mbox{sn}^3u=[\mbox{sn}(u,k)]^3$ etc.) are used.

For Eq.~(\ref{eq:redpoisson}) when the sum contains a single term
with $p=1$;
\begin{equation}
\frac{\mathrm d^2\varphi}{\mathrm dt^2}=
\frac{\varphi}{4}-2B\varphi^3
\label{eq:p1}
\end{equation}
where $B=2\pi D$, we first note that it does not involves the
independent variable explicitly, and therefore that, if $\varphi(t)$
is its solution, then $\varphi(t+c)$ where $c$ is an arbitrary
constant is also the solution. That is, one of the integration
constants of the general solution of Eq.~(\ref{eq:p1}) is related
to arbitrary translation of a particular solution. Next, let us think
of linear transformations of variables, $\varphi=
\lambda\tilde\varphi$ and $\tilde t=\alpha(t-t_0)$ where $t_0$ is an
arbitrary integration constant while $\lambda$ and $\alpha$ are
constants to be specified. With these new variables,
Eq.~(\ref{eq:p1}) is transformed to
\begin{equation}
\frac{\mathrm d^2\tilde\varphi}{\mathrm d\tilde t^2}=
\frac{1}{4\alpha^2}\tilde\varphi-2\frac{B\lambda^2}{\alpha^2}\tilde\varphi^3,
\end{equation}
which can reduce to any of the differential equations in
Eq.~(\ref{eq:deqel}) by proper choices of the constants. Note that
there are two constants, $\lambda$ and $\alpha$, to be specified but
Eq.~(\ref{eq:deqel}) contains only one parameter $k^2$. This
leaves one arbitrary degree of freedom, which essentially provides with
the remaining constant of integration. Then, we find the general
solutions of Eq.~(\ref{eq:p1}):
\begin{equation}
\varphi=\pm\left(\frac{1+c}{8B}\right)^{1/2}\
\mbox{cn}\left[\alpha_+(t-t_0),k_+\right]\,;\qquad
\varphi=\pm\left(\frac{4A}{c}\right)^{1/2}\
\mbox{sd}\left[\alpha_+(t-t_0),k_+\right]\,;
\end{equation}
\begin{equation}
\varphi=\pm\left(-\frac{8A}{1+c}\right)^{1/2}\
\mbox{nc}\left[\alpha_+(t-t_0),k_+\right]\,;\qquad
\varphi=\pm\left(-\frac{c}{4B}\right)^{1/2}\
\mbox{ds}\left[\alpha_+(t-t_0),k_+\right]\,;
\end{equation}
\begin{equation}
\varphi=\pm\left(\frac{1+c}{8B}\right)^{1/2}\
\mbox{dn}\left[\alpha_-(t-t_0),k_-\right]\,;\qquad
\varphi=\pm\left(-\frac{8A}{1+c}\right)^{1/2}\
\mbox{nd}\left[\alpha_-(t-t_0),k_-\right]\,;
\end{equation}
\begin{equation}
\varphi=\pm\left(\frac{8A}{1+c}\right)^{1/2}\
\mbox{sc}\left[\alpha_-(t-t_0),k_-\right]\,;\qquad
\varphi=\pm\left(-\frac{1+c}{8B}\right)^{1/2}\
\mbox{cs}\left[\alpha_-(t-t_0),k_-\right]
\end{equation}
where
\begin{equation}
c=(1+64AB)^{1/2}\ge0\,;\qquad
\alpha_+^2=\frac{c}{4}\,,\qquad
\alpha_-^2=\frac{1+c}{8}\,;\qquad
k_+^2=\frac{1}{2}\left(1+\frac{1}{c}\right)\,\qquad
k_-=\frac{1}{k_+},
\end{equation}
and $A$ and $t_0$ are two integration constants. In particular, $A$ is
chosen to correspond to the constant in Eq.~(\ref{eq:modpoisson}).
Here, all solutions are equivalent to one another in a sense that one
can be transformed to another with a proper choice of the integration
constant (which is, in general, allowed to be complex).

We note that, in principle, the solutions for $p=1/2$ and $p=2$ can
also be written in terms of Jacobi elliptic functions. However,
typical reduction procedures involve more complicated transformations
as well as the determination of constants though the solutions of
cubic or quartic polynomials.

\subsection{Solutions expressible using elementary functions}

The $\wp$-function reduces to elementary functions if the cubic
equation $f(z)=4z^3-g_2z-g_3=0$ has a degenerate solution or
equivalently $g_2^3=27g_3^2$. If the degenerate solution is zero, then
$g_2=g_3=0$ and $\wp(z;0,0)=z^{-2}$. If $e\ne0$ is the nonzero degenerate
solution of the cubic equation, since the quadratic coefficient of the
cubic polynomial $f(z)$ is null, the other solution of the cubic
equation is $-2e$ while $g_2=12e^2$ and $g_3=-8e^3$. Then, we find
that
\begin{equation}
\wp(z;12e^2,-8e^3)=
\left\{\begin{array}{lr}
3e\coth^2[(3e)^{1/2}z]-2e=
3e\,\mbox{csch}^2[(3e)^{1/2}z]+e\qquad&
\mbox{if }\ e>0\\
(-3e)\cot^2[(-3e)^{1/2}z]+(-2e)=
(-3e)\csc^2[(-3e)^{1/2}z]-(-e)\qquad&
\mbox{if }\ e<0
\end{array}\right.
\end{equation}
Similarly, Jacobi elliptic functions reduce to elementary functions
if $k=0$ or $k=1$. For example,
\begin{equation}
\begin{array}{lll}
\mbox{sn}(u,0)=\sin u;&
\mbox{cn}(u,0)=\cos u;&
\mbox{dn}(u,0)=1;\\
\mbox{sn}(u,1)=\tanh u;&
\mbox{cn}(u,1)=\mbox{sech}\,u;&
\mbox{dn}(u,1)=\mbox{sech}\,u.
\end{array}
\end{equation}

Using this, it is possible to find some nontrivial solutions of
Eq.~(\ref{eq:gle}) which can be written using elementary
functions for the case when the sum contains a single term with
$p=1/2$, 1, or 2. I) For $p=1/2$, they are
\begin{equation}
\psi=\frac{1}{6Br^{1/2}}\,;\qquad
\psi=\frac{1}{Br_0^{1/2}}\frac{r_0}{\left(r^{1/2}+r_0^{1/2}\right)^2}\,;
\qquad
\psi=-\frac{1}{Br_0^{1/2}}\frac{r_0}{\left(r^{1/2}-r_0^{1/2}\right)^2}\,;
\qquad
\psi=-\frac{1}{12Br^{1/2}}
\left\{1+3\tan^2\left[\frac{1}{4}\ln\left(\frac{r}{r_0}\right)\right]\right\}
\end{equation}
where $3B=8\pi D\ne0$. II) For $p=1$,
\begin{equation}
\psi=\pm\frac{1}{(8Br)^{1/2}}\,;\qquad
\psi=\pm\frac{1}{(Br_0)^{1/2}}\frac{r_0}{r+r_0}\,;\qquad
\mbox{for }\ B>0,
\end{equation}
\begin{equation}
\psi=\pm\frac{1}{(-8B)^{1/2}r^{1/2}}
\tan\left[\frac{1}{\sqrt{8}}\ln\left(\frac{r}{r_0}\right)\right]\,;\qquad
\psi=\pm\frac{1}{(-B)^{1/2}r_0^{1/2}}\frac{r_0}{r-r_0}\,;\qquad
\mbox{for }\ B<0,
\end{equation}
where $B=2\pi D$. III) For $p=2$,
\begin{equation}
\psi=\pm\frac{1}{(12B)^{1/4}r^{1/2}}\,;
\qquad
\psi=\pm\frac{1}{(3B)^{1/4}r^{1/2}}\left\{1+
3\tan^2\left[\frac{1}{2}\ln\left(\frac{r}{r_0}\right)\right]\right\}^{-1/2}\,;
\qquad
\psi=\pm\frac{1}{B^{1/4}r_0^{1/2}}
\frac{r_0}{\left(r^2+r_0^2\right)^{1/2}}\,;\qquad
\mbox{for }\ B>0,
\end{equation}
\begin{equation}
\psi=\pm\frac{1}{(-B)^{1/4}r_0^{1/2}}
\frac{r_0}{\left(r^2-r_0^2\right)^{1/2}}\,;\qquad
\psi=\pm\frac{1}{(-B)^{1/4}r_0^{1/2}}
\frac{r_0}{\left(-r^2+r_0^2\right)^{1/2}}\,;\qquad
\mbox{for }\ B<0,
\end{equation}
where $3B=4\pi D$. For all, $r_0>0$ is an arbitrary constant of
integration. In addition, if $B=0$, there are common solutions for
all $p$,
\begin{equation}
\psi=a+\frac{b}{r}\qquad
\mbox{for }\ B=0
\end{equation}
where $a$ and $b$ are arbitrary constants. These solutions include the
scale-free solution ($\psi\propto r^{-1/2}$ with the proportionality
coefficient uniquely determined by $p$) and the generalized Plummer
potentials of \citet{EA05}. With proper scaling, the oscillating
solution of $p=2$ case reduces the \citet{Sr62} solution of Lane-Emden
equation of index 5 in three dimension. However, except for the
generalized Plummer potential, none of these is physical.